\documentclass{aa}
\usepackage{graphicx}
\usepackage{epsfig}

\begin{document}

   \thesaurus{06     
              (13.25.1;  
               02.01.2;  
               08.02.3;  
               08.14.1)} 

\title{Optical observations of the black hole candidate
               XTE J1550-564 during the September/October 1998 outburst}
   \subtitle{}
\titlerunning{XTE J1550-564 during the Sep/Oct 1998 outburst}
  \authorrunning{S\'anchez-Fern\'andez et al.}

   \author{C. S\'anchez-Fern\'andez
          \inst{1}
   \and A. J. Castro-Tirado
          \inst{1,2}
   \and H. W. Duerbeck
          \inst{3}
   \and L. Mantegazza
          \inst{4}
   \and V. Beckmann 
          \inst{5}
   \and V. Burwitz 
          \inst{6}
   \and L. Vanzi 
          \inst{7}
   \and A. Bianchini
          \inst{8}
   \and M. Della Valle
          \inst{8}  
   \and A. Piemonte
          \inst{9}
   \and B. Dirsch 
          \inst{10}
   \and I. Hook 
          \inst{11}
   \and L. Yan
          \inst{12}
  \and A. Gim\'enez
          \inst{1,2}
          }
   \offprints{A. J. Castro-Tirado (ajct@iaa.es)}

   \institute{Laboratorio de Astrof\'{\i}sica Espacial y F\'{\i}sica 
              Fundamental (LAEFF-INTA), P.O. Box 50727,
              E-28080 Madrid, Spain.
   \and Consejo Superior de Investigaciones Cient\'{\i}ficas, 
        Instituto de Astrof\'{\i}sica de Andaluc\'{\i}a (IAA),
        P.O. Box 3004, E-18080 Granada, Spain.
   \and Free University Brussels (VUB), Pleinlaan 2, B-1050 Brussels, Belgium
   \and Osservatorio Astronomico di Brera, Via Bianchi 46, I-23807 
        Merate, Italy
   \and Hamburger Sternwarte, Universit\"at Hamburg, Gojenbergsweg 112, 
        D-21029 Hamburg, Germany. 
   \and Max-Planck-Institut fuer extraterrestische Physik, P.O. Box 1603
        D-85740 Garching, Germany.
   \and European Southern Observatory, Alonso de Cordova 3107, Santiago, Chile.
   \and Osservatorio Astronomico di Padova, vicolo dell'Osservatorio 5, 
         35122 Padova, Italy
   \and  Osservatorio Astrofisico di Asiago, Università di Padova, 
        I-36012 Asiago, Italy
   \and Sternwarte, University of Bonn, Auf dem H\"ugel 71 D-53121 Bonn,
        Germany.
   \and European Southern Observatory, Karl-Schwarzschild-Strasse 2-85748, 
        Garching bei M\"unchen, Germany. 
   \and Carnegie Observatories, 813 Santa Barbara St., Pasadena, 
        CA, 91101, USA. 
}

   \date{Received ; accepted}

   \maketitle

   \begin{abstract}
     
     We report here optical observations during the September/October 1998
     outburst of the black hole candidate XTE J1550-564. CCD photometry was
     obtained for the optical counterpart with the 0.9m Dutch telescope at
     La Silla since the onset of this event. We analysed 211 U, V and i
     frames, from Sep 10 to Oct 23. Stochastic flaring activity was
     observed on Sep 11-16, but no evidence for a superhump period, as seen
     in other soft X-ray transients, was found. An optical flare nearly 
     simultaneous to an X-ray flare that occurred on Sep 21 was observed in 
     the V-band. A reddened optical spectrum showed the typical emission lines
     corresponding to X-ray transients in outburst.  From the interstellar
     absorption lines, we derive E(B-V) = 0.70 $\pm$ 0.10 and suggest D
     $\approx$ 2.5 kpc. In such case, M$_{B}$ $\approx$ +7 mag. for the
     progenitor, which is consistent with the spectral type of a low-mass
     $\sim$ K0--K5 main-sequence companion.

      \keywords{Black hole physics --- Stars: neutron -- 
                Stars: binaries: general --- X-rays: general}
   \end{abstract}

%

\section{Introduction}

Low Mass X-ray Binaries (LMXBs) are systems formed by a low-mass companion
and a compact object. A subclass of LMXBs are the Soft X-ray Transients
(SXTs, so called X-ray Novae, although the physics is quite different from
the classical novae).  In these systems, sporadic outbursts are produced
due to some poorly understood mecha\-nism for which some mass is sporadically
transferred onto the compact primary via an accretion disk.  There are two
types of SXTs. In Type I, the compact object is a weakly magnetized neutron
star, whereas in Type II, the compact object is likely to be a black hole.
Normally they brighten in the course of a few days to become
one of the brightest sources in the X-ray sky, then declining in brightness
over the next few months. The X-ray spectra are often dominated by an
ultrasoft component and a hard X-ray tail.

The most recent Type II SXT, XTE J1550-564 was first detected on Sep 7.09
UT by the All-Sky Monitor on the Rossi X-Ray Timing Explorer.  First
detection yielded an intensity of $\sim$ 70 mCrab (2-12 keV; 5-$\sigma$
significance) (Smith and Remillard 1998).  The source
showed a steady rise to 1.7 Crab on Sep 15.  Thereafter, there was
increased variability, with the intensity reaching 3.2 Crab on Sep 18.7
UT.  On Sep 19 and 20, a large flare peaked at 6.8 Crab, and the intensity
fell back to the range of 2.7-3.6 Crab on Sep 20 and 21 (Remillard et al.
1998). The source was also detected in hard X-rays (20-100 keV) by the
Burst and Transient Source Experiment on the Compton Gamma-Ray Observatory
(Wilson et al. 1998).

An optical counterpart with V $\sim$ 16 mag. was proposed by Orosz et al. (1998) 
by means of V-band images obtained on Sep 8.99 UT. It is located at R.A. =
15$^h$50$^m$58$\fs$78, Decl.  = $-56\degr$ $28\arcmin$ $35\farcs0$ 
(equinox 2000.0)( $l^{II}$ = 326.2$\degr$, $b^{II}$ = -2.3$\degr$).
A likely radio counterpart to the X-ray transient, coincident with the
optical one, was detected on Sep 9 and 10 (Campbell-Wilson et al. 1998).

   \begin{figure}[tp]

\centering
\resizebox{8.2cm}{8.6cm}{\includegraphics{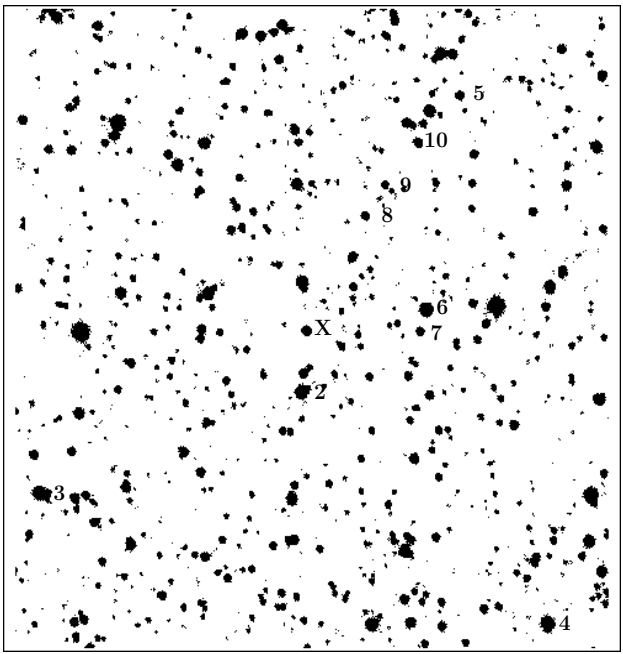}
\put(-250,247) {\bfseries {\Large X}}
\put(-250,197) {\bfseries {\Large 2}}
\put(-463,118) {\bfseries {\Large 3}}
\put(-50,17)   {\bfseries {\Large 4}}
\put(-120,430) {\bfseries {\Large 5}}
\put(-150,263) {\bfseries {\Large 6}}
\put(-155,243) {\bfseries {\Large 7}}
\put(-195,335) {\bfseries {\Large 8}}
\put(-180,358) {\bfseries {\Large 9}}
\put(-160,393) {\bfseries {\Large 10}}
}
        \caption[]{A $3\farcm77$ x $3\farcm77$ region centered at
          the position of the XTE J1550-564 optical counterpart, labelled
          as X. Local standard stars are numbered 2-10. North is up and
          East to the left.}
   \end{figure}

   \begin{table}
      \caption[t]{Local Standards in the XTE J1550-564 field.}
      \begin{flushleft}
      \begin{tabular}{lllccccc}
      \hline\noalign{\smallskip}
      {\tiny \#}&{\tiny $\alpha$(2000)}&{\tiny $\delta$(2000)}&{\tiny U}&{\tiny B}&{\tiny V}&{\tiny R}&{\tiny i}   \\
      \noalign{\smallskip}
      \hline\noalign{\smallskip}
{\tiny 2}&{\tiny 15~50~58.5}&{\tiny -56~28~57}&{\tiny 15.74}&{\tiny 15.44}&{\tiny 14.46}&{\tiny 13.98}&{\tiny 13.47}\\
{\tiny 3}&{\tiny 15~50~52.3} &{\tiny -56~29~40}&{\tiny 16.37}&{\tiny 15.65}&{\tiny 14.37}&{\tiny 13.67}&{\tiny 12.98}\\
{\tiny 4}&{\tiny 15~52~05.0} &{\tiny -56~30~15}&{\tiny 15.23}&{\tiny 13.78}&{\tiny 14.67}&{\tiny 13.40}&{\tiny 12.91}\\
{\tiny 5}&{\tiny 15~51~02.6} &{\tiny -56~27~04}&{\tiny 18.27}&{\tiny 17.70}&{\tiny 14.47}&{\tiny 15.87}&{\tiny 15.23}\\
{\tiny 6}&{\tiny 15~51~01.6} &{\tiny -56~28~23}&{\tiny 15.68}&{\tiny 15.22}&{\tiny 14.42}&{\tiny 14.02}&{\tiny 13.56}\\  
{\tiny 7}&{\tiny 15~51~01.7} &{\tiny -56~28~34}&{\tiny 18.51}&{\tiny 17.76}&{\tiny 16.66}&{\tiny 16.13}&{\tiny 15.55}\\
{\tiny 8}&{\tiny 15~51~00.0} &{\tiny -56~28~07}&{\tiny 18.53}&{\tiny 17.61}&{\tiny 16.28}&{\tiny 15.60}&{\tiny 14.90}\\
{\tiny 9}&{\tiny 15~51~00.5} &{\tiny -56~27~55}&{\tiny 18.56}&{\tiny 17.80}&{\tiny 16.60}&{\tiny 16.00}&{\tiny 15.36}\\
{\tiny 10}&{\tiny 15~51~01.4} &{\tiny -56~27~24}&{\tiny 17.89}&{\tiny 17.35}&{\tiny 16.23}&{\tiny 15.67}&{\tiny 15.08}\\
      \noalign{\smallskip}
      \hline
\noalign{\smallskip}
      \end{tabular}
 
     \end{flushleft}
      \end{table}

In this paper we present a series of optical observations of the XTE J1550-564 
counterpart taken with the 0.9m Dutch telescope at the
European Southern Observatory (ESO), La Silla between Sep 11 and Oct 23,
1998.


   \begin{figure}[tp]
      \epsfig{file=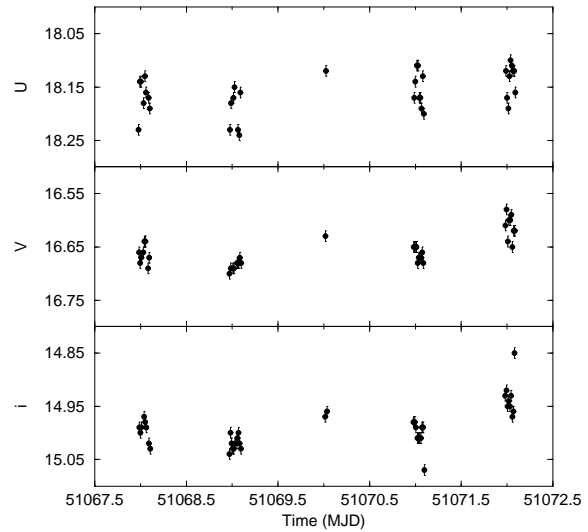, width=8.0cm, angle=-90.0} 
      \caption{Stochastic flaring activity in the  XTE J1550-564 
               light curve on Sep 11-16, prior to the strong
               flare that happened on Sep 19-20.}
   \end{figure}


\section{Observations}

Observations of the optical counterpart of XTE J1550-564 were carried out
with the 0.9m Dutch telescope at the ESO La Silla Observatory. 
Observations were taken between Sep 10 and Oct 23, 1998. The apparent 
proximity of the source to the Sun after Oct 23 prevented additional
optical observations.

The CCD used was a TEK CCD (512 x 512 pixels) that yielded a $3\farcm77$ x
$3\farcm77$ square field (see Figure 1). Typical exposure times were 600 s
for the Johnson-U filter, 180 s for the Johnson-V filter, and 120 s for the
Gunn-i filter per observing night.  On Sep 11-16, a series of exposures were
taken in the UVi-bands during 3 hr/night in order to search for short-term
variations.  The number of frames were 54 (U), 1 (B), 69 (V), 1(R) and
88 (i) during the observing pe\-riod.  The data were reduced using IRAF. The
images were processed to eliminate the electronic bias and flat field
corrected to remove the pixel-to-pixel sensitivity variations.  The optical
light curve was obtained when u\-sing the differences in magnitude between
the object and 9 field stars, given on Table 1. The typical uncertainties for 
the quoted coordinates and UBVRi magnitudes are 1$^{\prime\prime}$ and 0.01 
mag respectively. The magnitudes were calculated using the SExtractor software 
package (Bertin and Arnouts 1996).

A 300-s spectrum of the optical counterpart (range 3500-7400 \AA, 
resolution 2.0 \AA) was obtained with the 3.6-m ESO telescope 
(equipped with EFOSC2 and a B300 grism) on Sep 15.98 UT. He-Ar lamps
were used for wavelength calibration and the standard LTT 7379 for the
flux calibration. The reduction of the spectroscopic data was also carried 
out with IRAF. We corrected the spectrum for interstellar reddening 
following Cardelli et al. (1989), using  E(B-V) = 0.7, 
as discussed in the next section.


   \begin{figure}[tp]
\vspace{1cm}
      \epsfig{file=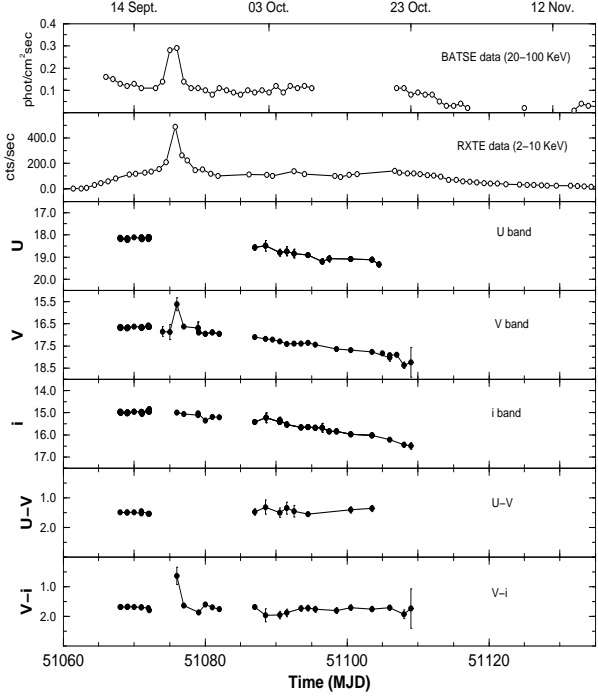,  height=9cm, width=9.5cm}        
         \caption{The light curve of XTE J1550-564 during the
          Sep/Oct 1998 outburst in $\gamma$-rays, X-rays (open
          circles) and UVi bands (filled circles). $\gamma$ and X-ray data
          are taken from BATSE and ASM public archives. 
          MJD is the modified Julian date, MJD = JD-2,400,000.5, 
          with Sep 1, 1998 being the MJD 51,056.}
    \end{figure}

%

    \begin{figure}[]
      \epsfig{file=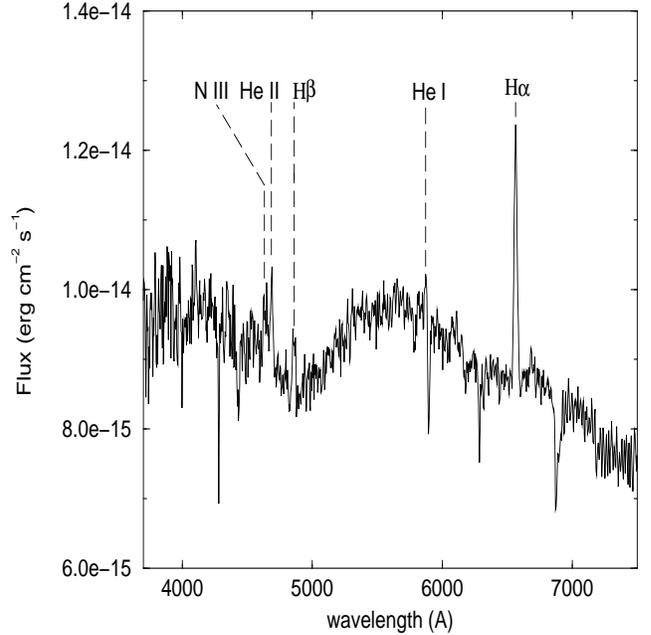, height=9.9cm, width=9.5cm}
      \caption{The dereddened optical spectrum of XTE J1550-564 taken on Sep 
        15.98 UT, following the prescription of Cardelli et al. (1989) for 
        E(B-V) = 0.7. 
        The spectrum shows a strong and broad emission line arising 
        from H-$\alpha$ (6563 \AA), as well as broad and weaker emission 
        lines from H-$\beta$ (4861 \AA), He II (4686 \AA) and 
        He I (5876 \AA). N III (4640 \AA) is also marginally 
        detected.}
   \end{figure}


\section{Results and discussion}

The first CCD images were taken on Sep 11 (MJD 51066), and showed XTE
J1550-564 with magnitudes V = 16.66 $\pm$ 0.01, U = 18.14 $\pm$ 0.01 and i=
14.98 $\pm$ 0.01. The object was monitored on an almost daily basis in the
UVi filters.  

The flux dropped by $\sim$ 1.5 mag during our obser\-ving period.  It is
striking that at the early epochs (MJD 50165 to 50175), the optical light
curve seems to remain appro\-ximately on a constant level, with some
stochastic flaring activity (Figure 2), whereas the X-ray spectrum is very
hard, although is gradually softening. Figure 3 shows the complete 
optical light curve based on our measurements. X-ray light curves from RXTE 
and BATSE have been included in the figure for comparison. Just after the 
flaring activity, a large brightening, with an amplitude of 1 mag in the 
V-band is detected,
coincident with the X-ray flare during which the V flux of the system
reaches V = 15.6 mag. We do not find support for the delay of $\approx$ 1
day claimed by Jain et al. (1999).  Unfortunately, we could not get
observations in the U filter during the maximum, so we cannot confirm that
this peak is also present in the U band.  However, there is no signature of
this peak in the i band. After this flaring episode, the light curve shows
a constant decrease in brightness, that extends at least until our last
data taken on MJD 51110.  We measure the following decay rates: dU/dt=
0.041 $\pm$ 0.006 mag/day, dV\-/dt= 0.048 $\pm$ 0.004 mag/day and di/dt=0.048
$\pm$ 0.005 mag/day.

We have searched for a superhump periodicity as seen in other SXTs (see
Della Valle, Masetti and Bianchini 1998 and references therein) by means of
Fourier analysis of the data obtained during Sep 11-16 following
the prescription of Horne and Baliunas (1986) for unequally spaced time
series. No convincing periodicity was found in the 0.05-2 day interval.

   \begin{table}
      \caption[t]{Observed emission lines in the XTE J1550-564 spectrum.}
      \begin{flushleft}
      \begin{tabular}{lcccc}
      \hline\noalign{\smallskip}
      Line & $\lambda_{0}$ &    EW     &          Flux            &  FWHM  \\
      ID   & \AA\   & \AA\ & (erg cm$^{-2}$ s$^{-1}$) & (km s$^{-1}$)   \\
      \noalign{\smallskip}
      \hline\noalign{\smallskip}
       N III      &  4640   &   4.3 & 2.8 $\times$ 10$^{-15}$ & 2900 \\
       He II      &  4686   &   3.2 & 2.6 $\times$ 10$^{-15}$ & 1550 \\
       H-$\beta$  &  4861   &   2.3 & 2.1 $\times$ 10$^{-15}$ & 1650 \\
       He I       &  5876   &   1.5 & 2.1 $\times$ 10$^{-15}$ &  900 \\
       H-$\alpha$ &  6563   &  11.9 & 2.0 $\times$ 10$^{-14}$ & 1250 \\
      \noalign{\smallskip}
      \hline
      \end{tabular}
      \end{flushleft}
    \end{table}

The optical spectrum shown in Figure 4 shows a strong and broad H-$\alpha$
emission line, and broad and weaker emission lines from H-$\beta$ and He II
(4686 \AA), as initially reported by Castro-Tirado et al. (1998).  He I
(5876 \AA) is strong and N III (4640 \AA) is also marginally detected. Line
characteristics are given in Table 2.  These lines are typical for soft
X-ray transients in outburst (Bradt and Mc\-Clintock 1983). The emission
lines and several interstellar absorption lines are superposed on a red
continuum.  The main interstellar features are the diffuse interstellar
band at 4430 \AA\ (EW = 1.9 \AA), the blend (EW = 2.4 \AA) due to the Na D
lines at 5890 and 5898 \AA\ and the blend due to the 6269 and 6282 \AA\
lines (EW = 1.7 \rm \AA). The `warp' feature around 5000 \AA\ has been 
observed in other SXTs at maximum or during the decline phase (Della Valle 
et al. 1991, Bianchini et al. 1997, Masetti et al. 1997) where the Balmer 
lines (especially H$_{\beta}$) were normally seen in emission filling in 
shallow absorptions.

The colour excess can be estimated on several ways. From the equivalent
width (EW) of the 4430 \AA\ band (EW = 1.9 \AA), we estimate a colour excess
of E(B-V) = 0.80 (Herbig 1975). From the empirical relation between the
width of the Na D lines given by Barbon et al. (1990) and considering EW =
2.4 \AA\ for the Na D lines, E(B-V) = 0.60.  Hereafter we will adopt
$<$E(B-V)$>$ = 0.70 $\pm$ 0.10, i.e. A$_{V}$ = 2.2 following the
relationship given by Savage and Mathis (1979).

The distance to RXTE J1550-564 could be estimated on the basis of the
linear relation between the equivalent width of the Na D lines and the 
distance (following Charles et al. 1989). We get D $\approx$ 2.5 kpc. 
From the work of Neckel and Klare (1980) it can be roughly seen how the 
absorption A$_{V}$ varies with the distance. On two fields 2$\degr$ 
away from the RXTE J1550-564 position, A$_{V}$ apparently remains constant up
to 2-3 kpc so no firm conclusion can be drawn. Hererafter we consider
D $\approx$ 2.5 kpc, that is significantly
different from D $\approx$ 6 kpc (Sobczak et al. 1999), a value obtained 
on the basis of the similarities of the optical and X-ray brightness of 
XTE J1550-564 with respect to the X-ray Nova Oph 1977. Given the large 
uncertainty in all the parameters, this distance estimate cannot be 
excluded.

The peak luminosity during the flare on MJD 51075 is, following Sobczak
et al. (1999), L = 2.0 $\times$ 10$^{38}$ (D/2.5 kpc)$^{2}$ erg s$^{-1}$,
which corresponds to the Eddington luminosity for M = 1.5 M$_{\odot}$ at 
2.5 kpc.  The optical to X--ray ratio during the flare is 
L$_{(V-band)}$/L$_{(2-10~\rm keV)}$ $\sim$ 2200, higher than the average value
of $\sim$ 500 found by van Paradijs \& McClintock (1995). 

For a quiescent magnitude of B $\approx$ 22 (Jain et al. 1999), i.e.
B $\approx$ 19 after the dereddening correction and a distance 
D $\approx$ 2.5 kpc, we derive M$_{B}$ $\approx$ +7 for the progenitor, 
a\-ssuming that there is no contribution from the disk in quiescence. 
This value is consistent with the spectral type of a low-mass $\sim$ K0--K5 
main-sequence companion (Allen 1976), similar to other SXTs for which 
radial velocities studies have been performed. The low-mass companion
is also supported by the large magnitude range from quiescence to the
outburst ($\sim$ 5 mag).

\section{Conclusions}

  Both the overall optical and X-ray light curves for XTE J1550-564 during
the first two months since the onset of the source resemble the light curves of
other type II SXTs (Tanaka and Shibazaki 1996). The spectrum shows emi\-ssion 
lines arising from H-$\alpha$, H-$\beta$, He II and N III, typical of soft 
X-ray transients in outburst.
From the interstellar absortion lines, we derive E(B-V) = 0.70 $\pm$ 0.10
and D $\approx$ 2.5 kpc, and M$_{B}$ $\approx$ +7 for the 
progenitor, which is consistent with a low-mass 
$\sim$ K0--K5 main-sequence companion. 
Only spectroscopic observations, to be performed when the system returns 
to quiescence, may lead to determine of the mass function of the system.
This will make po\-ssible to discern whether the compact
object in the XTE J1550-564 system is a neutron star or a black hole.

\begin{acknowledgements}
  We thank the referee, R. A. Remillard, for useful suggestions.
  One of us (CSF) is very grateful to J. Go\-rosabel for his valuable help.
  This work has been partially su\-pported by the spanish INTA grant 
  {\it Rafael Calvo Rod\'es} and the spanish CICYT grant ESP95-0389-C02-02.
\end{acknowledgements}

\end{document}